\documentclass[11pt]{article}

\usepackage[utf8]{inputenc}
\usepackage[T1]{fontenc}

\usepackage{enumitem}
\usepackage{multirow}
\usepackage{lmodern}
\usepackage{amsmath,amssymb}
\usepackage{verbatim}
\usepackage{pdflscape}
\usepackage{rotating}
\usepackage{framed}
\usepackage[numbers]{natbib}
\usepackage{graphicx}
\usepackage{hyperref}
\usepackage{array} \usepackage{makecell}   
\newcolumntype{P}[1]{>{\raggedright\arraybackslash}p{#1}}
\newcommand{\li}{\textbullet\ }

\newcommand{\code}[1]{\texttt{#1}}
\newcommand{\fct}[1]{\texttt{#1()}}
\newcommand{\proglang}[1]{\textsf{#1}}
\newcommand{\pkg}[1]{\texttt{#1}}

\title{generalRSS: Sampling and Inference for Balanced and Unbalanced Ranked Set Sampling in R}
\author{
    Chul Moon\thanks{Department of Statistics and Data Science, Southern Methodist University, Dallas, TX, USA. \texttt{chulm@smu.edu}}
    \and
    Soohyun Ahn\thanks{Department of Mathematics, Ajou University, Suwon, Republic of Korea. \texttt{shahn@ajou.ac.kr}} 
}

\begin{document}
\maketitle

\begin{abstract}
    Ranked set sampling (RSS) is a stratified sampling method that improves efficiency over simple random sampling (SRS) by utilizing auxiliary information for ranking and stratification. While balanced RSS (BRSS) assumes equal allocation across strata, unbalanced RSS (URSS) allows unequal allocation, making it particularly effective for skewed distributions. The \pkg{generalRSS} package provides extensive tools for both BRSS and URSS, addressing limitations in existing RSS software that primarily focus on balanced designs. It supports RSS data generation, efficient sample allocation strategies for URSS, and statistical inference for both balanced and unbalanced designs. This paper presents the RSS methodology and demonstrates the utility of \pkg{generalRSS} through two medical data applications: a one-sample mean inference and a two-sample area under the curve (AUC) comparison using NHANES datasets. These applications illustrate the practical implementation of URSS and show how \pkg{generalRSS} facilitates ranked set sampling and inference in real-world data analysis.
\end{abstract}


\section[Introduction]{Introduction} \label{sec:intro}

Randomized experiments are fundamental to modern scientific discovery and typically depend on simple random sampling (SRS) to select units. While increasing the sample size can enhance the efficiency of such experiments, this approach may be impractical under resource constraints. Ranked set sampling (RSS), first introduced by \cite{mcintyre1952method}, offers a cost-effective alternative. RSS is a stratified sampling method that utilizes auxiliary ranking information to create strata \citep{StokesS:1988}. When each rank stratum includes an equal number of samples, it is called balanced RSS (BRSS). In contrast, unbalanced RSS (URSS) occurs when the number of samples differs across strata. 

RSS has a rich history of research and development \citep[][and references therein]{ChenBS:2006,Wolfe:2012}. Research has shown that BRSS consistently outperforms SRS with an equivalent sample size in estimation precision, as demonstrated by \cite{Takahashi:1968}. Numerous studies have explored the efficiency of RSS estimators \citep[]{dell1972ranked,david1972ranked,stokes1980inferences}. For skewed distributions, URSS estimators can achieve even greater efficiency than BRSS and SRS counterparts \citep{ahn2017unbalanced, Ahn:2024, bocci2010ranked, Chen:2000, Ozturk:2004}. However, the performance of URSS depends heavily on the allocation of replicates among strata. Improper allocation can lead to inefficiencies, sometimes even worse than SRS. To address this, several studies \citep{bhoj2020simple, Chen:2000,mcintyre1952method,Wang:2004} have proposed appropriate allocation rules, with Neyman allocation being the most widely used due to its optimal variance-minimizing properties when estimating the population mean. As a result, most of the existing literature on URSS has focused on this allocation strategy \citep{ChenBS:2006,Takahashi:1968,wang2017unbalanced}. Recently, \cite{Ahn:2022} defined a sufficient set of allocation schemes to ensure that URSS achieves greater efficiency in population mean estimation compared to BRSS. They also introduced two practical allocation adjustments to enhance the efficiency of URSS designs beyond that of BRSS.

Several \proglang{R} packages are available for RSS, including \pkg{RSSampling} \citep{RSSampling}, \pkg{NSM3} \citep{NSM3}, \pkg{RSStest} \citep{RSStest}, and \pkg{RankedSetSampling} \citep{RankedSetSampling}. \pkg{RSSampling} supports a wide range of RSS extensions, offering both sampling tools and statistical inference. \pkg{NSM3} includes only classical RSS as a sampling procedure and provides critical value calculations for nonparametric inference. \pkg{RSStest} focuses on sampling and mean testing for RSS and MRSS and includes simulation tools under normal distributions. Finally, \pkg{RankedSetSampling} extends its scope to joint probability sampling (JPS) and offers mean and variance estimation for RSS. However, these packages are generally limited to balanced RSS designs, restricting their applicability for both sampling and inference in scenarios that require more flexibility.

BRSS designs are commonly implemented due to their simplicity and ease of implementation. However, in real-world survey studies, missing data often results in URSS scenarios, where BRSS methodologies become impractical. URSS offers a flexible alternative by allowing unequal allocation across strata, making it particularly effective for skewed distributions where unequal allocation can significantly improve estimation efficiency. Despite its advantages, implementing URSS has been challenging due to the lack of accessible tools in existing software packages.

\pkg{generalRSS} package was developed to address these limitations by fully supporting URSS while maintaining compatibility with BRSS. It provides parametric and nonparametric inference tools for population means, medians, AUCs, and proportions. Additionally, it offers functions for generating ranked set samples with specified allocations and optimizing allocation strategies to improve the efficiency of URSS designs. The flexibility of URSS allows users to tailor sampling designs to real-world constraints, making it particularly valuable when balanced designs are impractical. By treating BRSS as a special case of URSS with equal allocation across strata, \pkg{generalRSS} extends the applicability of RSS beyond BRSS methods. This versatility makes it a practical tool for a wide range of applications, including environmental or medical studies, where flexible sampling strategies are essential.

The rest of this paper is organized as follows. Section 2 provides an overview of ranked set sampling, with Section 2.1 covering BRSS and Section 2.2 discussing URSS, including their respective procedures. Section 3 describes the analysis approaches and functions available in \pkg{generalRSS} and compares them with existing \proglang{R} packages. Section 4 evaluates the performance of RSS relative to SRS in two inference problems using data from the US National Health and Nutrition Examination Survey (NHANES) \citep{NHANES,centers2023national}. Finally, Section 5 summarizes the capabilities of \pkg{generalRSS}, its applications in inference problems, and future directions for inference methods and sampling designs.


\section{Ranked Set Sampling} \label{sec:rss}

In this section, we illustrate the sampling procedure of RSS by detailing its design and associated notations. An RSS dataset of size $n$ can be represented as 
\begin{equation}\label{eqn:rss}
  \{(y_i, h_i, r_i), i=1,2,\cdots, n\},  
\end{equation}
where $y_i$ denotes the $i$-th observation, $h_i$ represents its rank among $r_i$ independent observations, and $r_i$ denotes the set size. Typically, RSS assumes a fixed set size, i.e., $r_i=H$ for all $i=1,2,\cdots,n$, and $h_i$ takes values in $\{1,\cdots, H\}$. The number of observations assigned to rank $h$ is given by $n_h=\sum_{i=1}^H I(h_i=h)$, where $I(\cdot)$ denotes the indicator function. The total sample size of the RSS data is then given by $n=\sum_{h=1}^H n_h$. These notations define the structure of RSS data and its implementation in \pkg{generalRSS}. 

RSS data can also be represented using the notation $Y_{[h],r}$ denoting the $r$-th observation assigned to rank $h$, with $h=1,\cdots,H$ and $r=1,\cdots,n_h$. The RSS sampling process consists of the following steps:\\

\noindent Step 1: Select a simple random sample of size $H$ from the target population.\\
Step 2: Rank the $H$ sampled units using an auxiliary variable without measuring the variable of primary interest.\\
Step 3: Measure the unit ranked as the $h$-th smallest and discard the remaining units.\\
Step 4: Repeat Steps 1-3 for each rank $h$ up to $H$.\\

\begin{figure}[!ht]
    \centering
    \includegraphics[width=1\linewidth]{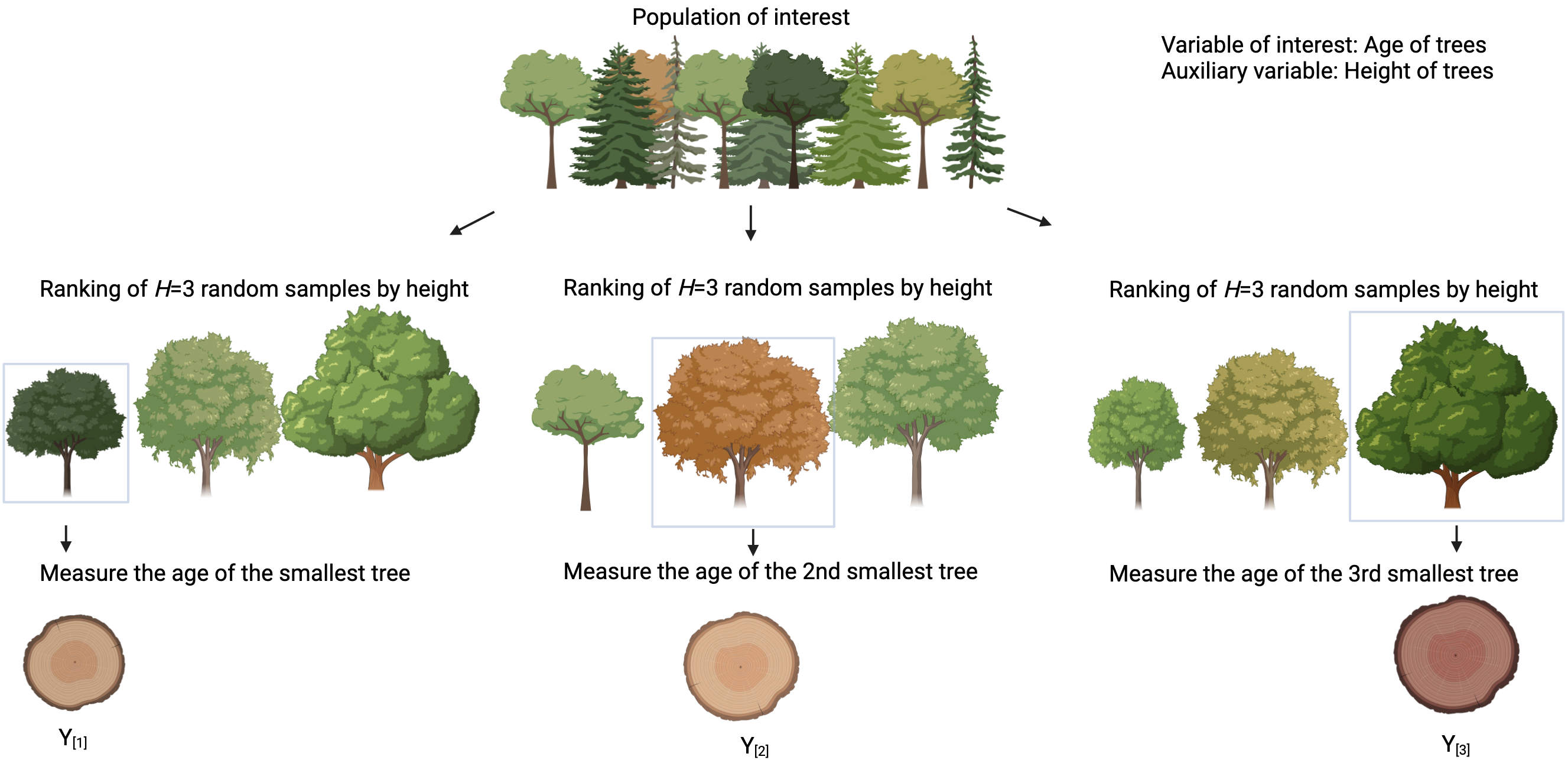}
    \caption{Example of a cycle of RSS of a set size of $H=3$ and ranked set samples. The variable of interest (Y) is tree age and the auxiliary variable (X) is tree height.}
    \label{fig:cycle}
\end{figure}

\noindent These steps define one cycle of RSS, as illustrated in Figure~\ref{fig:cycle}. In each cycle, a SRS of size $H$ is selected, ranked, and one unit is measured per stratum, yielding a sample consisting of $H$ strata with one observed unit. By repeating these cycles, multiple observations are collected within each stratum, ultimately forming the final ranked set sample. 

\subsection{Balanced RSS}

BRSS assumes equal allocation across strata, meaning that each rank $h$ is assigned the same number of observations, i.e., $n_h=m$ for every $h=1,2,\cdots, H$. To obtain this, the RSS sampling cycle is repeated $m$ times, resulting in a BRSS design with the sample allocation $(m, m, \cdots, m)$. For $H=3$, a BRSS procedure with $m$ cycles can be represented as: 
\[
\begin{aligned}
\textbf{Cycle $r=1$} \\
{\bf X}_{[1]11} \leq X_{[2]11} \leq X_{[3]11} &\implies Y_{[1]1} \\
X_{[1]21} \leq {\bf X}_{[2]21} \leq X_{[3]21} &\implies Y_{[2]1} \\
X_{[1]31} \leq X_{[2]31} \leq {\bf X}_{[3]31} &\implies Y_{[3]1} \\
\\
\textbf{Cycle $r=2$} \\
{\bf X}_{[1]12} \leq X_{[2]12} \leq X_{[3]12} &\implies Y_{[1]2} \\
X_{[1]22} \leq {\bf X}_{[2]22} \leq X_{[3]22} &\implies Y_{[2]2} \\
X_{[1]32} \leq X_{[2]32} \leq {\bf X}_{[3]32} &\implies Y_{[3]2} \\
\\
\cdots \cdots \cdots \cdots \cdots \cdots \cdots \cdots & \cdots \cdots \cdots
\\
\textbf{Cycle $r=m$} \\
{\bf X}_{[1]1m} \leq X_{[2]1m} \leq X_{[3]1m} &\implies Y_{[1]m} \\
X_{[1]2m} \leq {\bf X}_{[2]2m} \leq X_{[3]2m} &\implies Y_{[2]m} \\
X_{[1]3m} \leq X_{[2]3m} \leq {\bf X}_{[3]3m} &\implies Y_{[3]m} \\
\end{aligned}
\]
where $X_{[h]ir}$ denotes the $h$-th smallest auxiliary variable of the $i$-th SRS of size $H$, selected in the $r$-th cycle.
The resulting BRSS dataset is given by:
\[\{Y_{[1],1},\cdots, Y_{[1],m}, Y_{[2],1}, \cdots, Y_{[2],m}, \cdots, Y_{[H],1},\cdots, Y_{[H],m}\}.\]

\subsection{Unbalanced RSS}

While BRSS assumes equal allocation across strata, real-world applications often involve scenarios where such balance is impractical. For instance, skewed data, missing observations, or resource constraints can result in an unbalanced design. URSS allows for greater flexibility by permitting unequal allocations across strata, enabling more efficient sampling when strata exhibit higher variability or measurement difficulty. For example, if the population of interest has strata with differing levels of variability in measurements, allocating more samples to strata with higher variability can improve estimation efficiency, which BRSS cannot achieve. 

Unlike BRSS, which completes a full cycle by selecting one sample for each stratum in every iteration, URSS follows an incomplete cycle structure, where strata exit the sampling process once they reach their allocated sample sizes $n_h$. This means that different strata complete different numbers of iterations, leading to an unbalanced design. In Step 4 of the URSS procedure, Steps 1-3 are repeated only for strata that have not yet reached $n_h$, ensuring that each stratum meets its predefined sample allocation.

For instance, when $H=3$, consider an allocation where $n_1 < n_2 < n_3$. Then, a URSS with sample allocation ($n_1, n_2, n_3$) is obtained as:
\[
\begin{aligned}
\textbf{Cycle $r=1,\cdots,$} &\textbf{$n_1$} \\
{\bf X}_{[1]1r} \leq X_{[2]1r} \leq X_{[3]1r} &\implies Y_{[1]r} \\
X_{[1]2r} \leq {\bf X}_{[2]2r} \leq X_{[3]2r} &\implies Y_{[2]r} \\
X_{[1]3r} \leq X_{[2]3r} \leq {\bf X}_{[3]3r} &\implies Y_{[3]r} \\
\\
\textbf{Cycle $r=n_1+1,\cdots,$} & \textbf{$n_2$} \\
X_{[1]2r} \leq {\bf X}_{[2]2r} \leq X_{[3]2r} &\implies Y_{[2]r} \\
X_{[1]3r} \leq X_{[2]3r} \leq {\bf X}_{[3]3r} &\implies Y_{[3]r} \\
\\
\textbf{Cycle $r=n_2+1,\cdots,$} &\textbf{$n_3$} \\
X_{[1]3r} \leq X_{[2]3r} \leq {\bf X}_{[3]3r} &\implies Y_{[3]r} \\
\end{aligned}
\]

For a general $H$, the resulting URSS data can be represented as:
\[\{Y_{[1],1},\cdots, Y_{[1],n_1}, Y_{[2],1}, \cdots, Y_{[2],n_2}, \cdots, Y_{[H],1},\cdots, Y_{[H],n_H}\}.\]
In practice, missing values during the sampling process in BRSS often lead to a URSS design. This highlights the importance of methodologies that can accommodate both balanced and unbalanced designs seamlessly.

\section[generalRSS package]{\pkg{generalRSS} package}

In Section~\ref{sec:rss}, we introduced the RSS procedure, which involves selecting samples from a population and ranking them based on an auxiliary variable. In this section, we demonstrate how the \pkg{generalRSS} package facilitates the implementation of RSS methodologies, enabling efficient sampling and accurate inference of population parameters.

The \pkg{generalRSS} package, available on CRAN at \url{https://cran.r-project.org/package=generalRSS}, can be installed and loaded using the following commands:
\begin{verbatim}
R> install.packages("generalRSS")
R> library("generalRSS")
\end{verbatim}
The package incorporates some functions from \proglang{R} packages \pkg{emplik} \citep{emplik} and \pkg{rootSolve} \citep{rootSolve} for solving empirical likelihood problems. It is designed to address both sampling and inference problems in RSS with a focus on unbalanced designs. The package comprises two main components: (i) functions for sampling and sample allocation and (ii) functions for statistical inference.

\subsection{Sampling and Allocation Design}
The \pkg{generalRSS} package introduces two sampling functions to facilitate RSS procedures as well as a design function to calculate efficient sample allocations for URSS as summarized in Table~\ref{tab:sampling}.
\begin{table}[!ht]
    \centering
    \resizebox{\textwidth}{!}{
    \begin{tabular}{l|l} \hline
        Function & Description \\ \hline
        \fct{rss.sampling} & Generate ranked set samples \\
        \fct{rss.simulation} & Generate example ranked set samples\\
        \fct{rss.prop.sampling} & Generate ranked set samples for proportions \\
        \fct{rss.prop.simulation} & Generate example ranked set samples for proportions \\
        \fct{rss.design} & Calculate efficient sample allocations for RSS \\ \hline
    \end{tabular}
    }
    \caption{Functions for the sampling and allocation design in the \pkg{generalRSS} package.}
    \label{tab:sampling}
\end{table}
The function \fct{rss.sampling} generates ranked set samples directly from a population dataset containing an auxiliary variable ($X$) and, optionally, a variable of interest ($Y$). 
When the auxiliary variable coincides with the variable of interest (e.g., $X=Y$), perfect ranking is assumed.
If the variable of interest ($Y$) is not provided, the function selects sample IDs based on $X$, returning a data frame containing IDs and ranks ($r_i$). If $Y$ is provided, the function directly returns RSS data with observation ($y_i$). For illustration, we use the \code{iris} dataset as the population, treating \code{Sepal.Length} as the outcome variable ($Y$) and \code{Petal.Length} as the auxiliary variable ($X$), with a set size $H=3$ and sample allocations $(n_1=2, n_2=2, n_3=2)$:
\begin{verbatim}
R> data("iris")
R> id = 1:nrow(iris)
R> rss.data = rss.sampling(ID=id, Y=NULL, X=iris$Petal.Length,
+  H=3, nsamp=c(2,2,2))
R> head(rss.data)
  rank  ID
1    1  14
2    1  50
3    2 143
4    2  92
5    3 137
6    3 148
\end{verbatim}
Since $Y$ was not provided, the function returns only the selected IDs and ranks. The actual measurement of $Y$ must be collected for these selected samples to complete the dataset. 
When $Y$ is included in the population dataset, the function automatically extracts the corresponding values and returns a data frame containing ranks ($r_i$), IDs, and observations ($y_i$).
\begin{verbatim}
R> rss.data = rss.sampling(ID=id, Y=iris$Sepal.Length, 
+  X=iris$Petal.Length, H=3, nsamp=c(2,2,2))
R> head(rss.data)
  rank  ID   y
1    1  14 4.3
2    1  50 5.0
3    2 143 5.8
4    2  92 6.1
5    3 137 6.3
6    3 148 6.5
\end{verbatim}
The \code{nsamp} parameter specifies the number of cycles ($n_h$) for each stratum. When $n_h=m$ for all $h$, the output is BRSS data. If $n_h$ varies across strata, the design corresponds to a URSS scheme.

The function \fct{rss.simulation} generates ranked set samples by simulating data from selected probability distributions, including normal, t, and lognormal distributions, for illustrative purposes. This function allows users to specify RSS design parameters, such as the set size ($H$) and the number of cycles ($n_h$), to customize the simulation for different scenarios. Additionally, the function supports adjustable mean shifts through the \code{delta} parameter, enabling the simulation of populations with group-specific differences. 

The ranking accuracy in the simulated data is controlled through a linear ranking model defined as $X_i=Y_i+\epsilon_i$ for $i=1,2,\cdots,n$, where $\epsilon_i$ represents independent normal random variables with mean 0 and variance chosen to achieve a specific correlation $\rho=Corr(X, Y)$. The \code{rho} parameter determines the correlation between the outcome variable ($Y$) and the auxiliary variable ($X$). A value of $\rho=1$ corresponds to perfect ranking (i.e., no ranking errors), while values between 0 and 1 represent imperfect ranking, where ranking errors increase as $\rho$ decreases. The function returns a data frame containing ranks ($r_i$) and outcomes ($y_i$). For example, the following code generates RSS data from a normal distribution with a mean shift of $\delta=0$ and a ranking quality of $\rho=0.8$:
\begin{verbatim}
R> rss.simulation(H=3, nsamp=c(2,2,2), dist="normal", rho=0.8, 
+  delta=0)
\end{verbatim}    
The functions \fct{rss.prop.sampling} and \fct{rss.prop.simulation} are the sampling and simulation functions for proportions. The variable of interest $(Y)$ is a binary variable of 0 and 1, corresponding to failure and success, respectively. We assume the perfect ranking $(X=Y)$ for proportions. For illustration, we use the \code{iris} dataset as the population, defining both the outcome and the auxiliary variable ($X=Y$) as a binary indicator of whether \code{Sepal.Length} is less than 5.8:
\begin{verbatim}
data(iris)
id = 1:nrow(iris)
X = ifelse(iris$Sepal.Length < 5.8,0,1)
head(X)
0 0 0 0 0 0
\end{verbatim}
Once the binary variable is defined, we apply \fct{rss.prop.sampling} to generate RSS data with a set size of $H=3$ and sample allocations $(n_1=2, n_2=2, n_3=2)$:
\begin{verbatim}
rss.prop.data = rss.prop.sampling(ID=id, X=X, H=3,
+  nsamp=c(2,2,2))
rss.prop.data
  rank  ID
1    1  14
2    1  50
3    2 118
4    2  92
5    3 137
6    3 147
\end{verbatim}
The \fct{rss.prop.simulation} function generates ranked set samples for proportions by simulating data based on a true population proportion $p$. The \code{p} parameter represents the true proportion of successes and serves as the success probability in a binomial distribution of $Y$. For example, the following code generates RSS data with a proportion of $p=0.6$:
\begin{verbatim}
R> rss.prop.data = rss.prop.simulation(H=3, nsamp=c(2,2,2),
+  p=0.6)
R> rss.prop.data
  rank  y
1    1  0
2    1  0
3    2  1
4    2  1
5    3  1
6    3  1
\end{verbatim}
The \fct{rss.design} function evaluates the efficiency of a current URSS design for mean estimation based on either an initial RSS dataset or predefined sample allocation. If the design is inefficient compared to SRS or BRSS, the function calculates improved sample allocations to enhance efficiency. The function offers three allocations for estimating a population mean: integer Neyman allocation \citep{wright2012equivalence}, adjusted Neyman allocation \citep{Ahn:2022}, and local ratio consistent (LRC) allocation \citep{Ahn:2022}. These methods ensure greater efficiency with minimal additional samples compared to the original URSS design. For example, we consider a simulated URSS dataset with an initial allocation of $(3,10,5)$. The function calculates optimized allocations as follows:
\begin{verbatim}
R> rss.data = rss.simulation(H=3, nsamp=c(3,10,5), dist="t",
+  rho=1, delta=0)
R> rss.design(rss.data)
$original.n
n1 n2 n3 
 3 10  5 

$Integer.Neyman
n1 n2 n3 
 4  5  9 

$Adj.Neyman
n1 n2 n3 
 4 10  9 

$LRC.allocation
n1 n2 n3 
 6 10 10 
\end{verbatim}
In this example, the function first computes the integer Neyman allocation ($n_1=4, n_2=5, n_3=9$) and then recommends adding 1 sample to the first stratum and 4 samples to the third stratum, resulting in an adjusted Neyman allocation ($n_1=4, n_2=10, n_3=9$). Similarly, it computes the LRC allocation ($n_1=6, n_2=10, n_3=10$) by adding 3 samples to the first stratum and 5 samples to the third stratum. These adjusted allocations improve the efficiency of the original URSS design.

The \fct{rss.design} function also computes the optimal Neyman allocation for estimating a population proportion under perfect rankings \citep{chen2006unbalanced}. The following example compares an initial URSS allocation $(n_1=10,n_2=15,n_3=20)$ to the computed Neyman allocation:
\begin{verbatim}
R> rss.prop.data = rss.prop.simulation(H=3, nsamp=c(10,15,20),
+  p=0.5)
R> rss.design(rss.prop.data, prop=TRUE)
$original.n
n1 n2 n3 
10 15 20 

$Neyman.proportion
      n1       n2       n3 
12.45017 19.36741 13.18242 
\end{verbatim}
Here, the computed Neyman allocation aligns more closely to the true Neyman allocation $(n_1=12.814, n_2=19.373, n_3=12.814)$, as suggested by \cite{chen2006unbalanced}, compared to the initial URSS design.

\subsection{Statistical Inference}

The \pkg{generalRSS} provides six statistical inference functions for estimating and testing the population means, medians, proportions, and AUCs using RSS data. These functions include both parametric and nonparametric methods, as summarized in Table~\ref{tab:inference}.

\begin{table}[!ht]
    \centering
    \resizebox{\textwidth}{!}{
    \begin{tabular}{c|c} \hline 
        Function & Description \\ \hline
        \fct{rss.z.test} & RSS z-test for one-sample and two-sample problems \\
        \fct{rss.t.test} & RSS t-test for one-sample and two-sample problems\\
        \fct{rss.ELR.test} & RSS empirical likelihood ratio test for one-sample problems \\
        \fct{rss.sign.test} & RSS Sign test for one-sample problems \\
        \fct{rss.prop.test} & RSS proportion test for one-sample problems \\
        \fct{rss.AUC.test} & RSS empirical likelihood ratio test for two-sample problems \\ \hline
    \end{tabular}
    }\caption{Statistical inference functions in the \pkg{generalRSS} package.}
    \label{tab:inference}
\end{table}

The \fct{rss.z.test} function provides point estimation, confidence intervals (CIs), and hypothesis testing for the population mean using a normal approximation for RSS data \citep{ChenBS:2006, Ahn:2024}. 
It uses the asymptotic pivotal method to test $H_0:\mu=\mu_0$, where the test statistic is:
\begin{equation}
    \frac{\widehat{\mu}_{\rm RSS}-\mu_{0}}{\widehat{\sigma}_{\widehat{\mu}_{\rm RSS}}} \overset{d}{\longrightarrow}N(0,1)
    \label{eq:pivot}
\end{equation}
where $\widehat{\mu}_{\rm RSS}=\frac{1}{H}\sum_{h=1}^H \frac{1}{n_h}\sum_{r=1}^{n_h} Y_{[h],r}$, $\widehat{\sigma}^2_{\widehat{\mu}_{\rm RSS}}=\frac{1}{H^2}\sum_{h=1}^H\frac{1}{n_h(n_h-1)}\sum_{r=1}^{n_h}\left(Y_{[h],r}-\bar{Y}_{[h]}\right)^2$, and $\bar{Y}_{[h]}=\sum_{r=1}^{n_h} Y_{[h],r}/n_{h}$.

Similarly, the \fct{rss.t.test} function performs mean inference using a t-distribution approximation of the pivot statistics in equation~(\ref{eq:pivot}) \citep{Ahn:2014}. It supports two methods for estimating degrees of freedom, \code{sample} and \code{naive}, which are controlled by the \code{method} parameter. 
Both \fct{rss.z.test} and \fct{rss.t.test} functions handle one-sample and two-sample problems using the same pivotal statistics but differ in their approximation methods. The following example demonstrates the \fct{rss.t.test} function for a two-sample problem, testing the hypothesis that the population mean difference is 0:
\begin{verbatim}
R> rss.data1 = rss.simulation(H=3, nsamp=c(6,6,6), dist="normal",
+  rho=0.8, delta=0)
R> rss.data2 = rss.simulation(H=3, nsamp=c(6,8,10), dist="normal", 
+  rho=0.8, delta=0.5)
R> rss.t.test(data1=rss.data1, data2=rss.data2, alpha=0.05,
+  alternative="two.sided", mu0=0, method="naive")
$RSS_mean
[1] -0.05269583  0.59074771

$CI
[1] -0.9032096 -0.3836775

$t
[1] -5.023613

$df
[1] 36

$p.value
[1] 1.397776e-05
\end{verbatim}
The \fct{rss.ELR.test} function provides a semi-parametric one-sample mean test using an empirical likelihood ratio (ELR) test \citep{Ahn:2024}. This method offers a flexible alternative to traditional parametric approaches, particularly when normality assumptions are not appropriate. \fct{rss.ELR.test} uses the empirical likelihood function which is defined as:
\begin{equation}\nonumber
L(\mu_0)=\sup \left\{\prod_{h=1}^H\prod_{r=1}^{n_h}  p_{h,r}: \sum_{r=1}^{n_h} p_{h,r}=\frac{1}{H} ~~\mbox{and}~~ \sum_{r=1}^{n_h} p_{h,r} Y_{[h],r}=\mu_0 \right\},
\end{equation} 
where $p_{h,r}$ is a mass probability on $Y_{]h],r}$.
The following example demonstrates how to test the hypothesis $H_0: \mu=0$ using \fct{rss.ELR.test}:
\begin{verbatim}
R> rss.ELR.test(data=rss.data1, alpha=0.05, mu0=0)
$RSS_mean
[1] -0.05269583

$CI
[1] -0.3153479  0.2204137

$`-2*log.LR`
[1] 0.1488371

$p.value
[1] 0.6996491
\end{verbatim}
The \fct{rss.sign.test} function performs a nonparametric one-sample sign test, providing median estimation, CIs, and hypothesis testing. For BRSS, the function follows the method of \cite{hettmansperger1995ranked} whereas it implements the approach of \cite{Barabesi:2001} for URSS. Under the null hypothesis $H_0:M=M_0$, the test statistic for BRSS is:
\begin{equation} \nonumber 
n^{-1/2}\left(S_{{\rm RSS}}^{+}-\frac{n}{2}\right)
 \overset{d}{\longrightarrow}N\left(0,~\frac{\eta^{2}}{4}\right),
\end{equation} 
where $S_{{\rm RSS}}^{+}=\sum_{i=1}^{n}{\rm I}\big( y_{i}-M_{0}>0\big)$, $\eta^{2}=1-\frac{4}{H}\sum_{h=1}^{H}\left\{ B\left(h,H-h+1,\frac{1}{2}\right)-\frac{1}{2}\right\} ^{2}$, and $B(h,s,q)$ is the cumulative distribution function of the beta distribution with parameters $h$ and $s$ for $0 \le q \le 1$. For URSS, the test statistic follows:
\begin{equation} \nonumber
\left(S_{{\rm RSS}}^{+}- \sum_{h=1}^H n_h (1-\beta_h) \right)
 \overset{d}{\longrightarrow}N\left(0,\sum_{h=1}^H n_h \beta_h (1-\beta_h)\right),
\end{equation}
where $\beta_h= B\left(h, H-h+1,\frac{1}{2}\right)$. The following example demonstrates how to test whether the population median is 0 using \fct{rss.sign.test}:
\begin{verbatim}
R> rss.sign.test(data=rss.data1, alpha=0.05, 
+  alternative="two.sided", median0=0)
$RSS_median
[1] -0.09226903

$sign
[1] 12

$CI
[1] -0.3059627  0.3796395

$z
[1] -0.7745967

$p.value
[1] 0.438578
\end{verbatim}
The \fct{rss.prop.test} function offers proportion estimation, CIs, and hypothesis testing for binary outcome variables ($Y$). The unbiased RSS estimator for $p$ is $\hat{p}=\frac{1}{H}\sum_h=1^H\frac{1}{n_h}\sum_{r=1}^{n_h}Y_{[h]r}$ and its variance is $\text{Var}(\hat{p})=\frac{1}{H^2}\sum_{h=1}^{H}\frac{1}{n_h}\hat{p}(1-\hat{p})$. Under perfect rankings, we can use a normal approximation for proportions of RSS data \citep{chen2006unbalanced,ahn2022continuity}:
\begin{equation} \nonumber 
\hat{p}-p \overset{d}{\longrightarrow}N\left(0,\frac{1}{H} \sum_{l=H-h+1}^H \binom{H}{l}p^l(1-p)^{H-l} \right).
\end{equation} 
We extend the CI method proposed by \cite{zamanzade2020using} for BRSS to URSS by $\hat{p} \pm z_{1-\frac{\alpha}{2}}\sqrt{\frac{1}{H^2}\sum_{h=1}^H \frac{1}{n_h}\hat{p}_h(1-\hat{p}_h)}$ where $\hat{p}_h=\sum_{l=H-h+1}^H \binom{H}{l}\hat{p}^l(1-\hat{p})^{H-l}$. The following example illustrates how to use \fct{rss.prop.test} to test the null hypothesis $H_0: p=0.2$ when the true population proportion is $p=0.6$.
\begin{verbatim}
R> rss.prop.data = rss.prop.simulation(H=3, nsamp=c(12,9,6), 
+  p=0.6)
R> rss.prop.test(data=rss.prop.data, alpha=0.05, 
+  alternative="two.sided", p0=0.2)
$RSS_prop
[1] 0.4907407

$CI
[1] 0.3367646 0.6447169

$pstat
[1] 3.700841

$p.value
[1] 0.0002148859
\end{verbatim}
Additionally, the \fct{rss.AUC.test} function conducts a semi-parametric ELR test for comparing the AUC between two groups based on RSS data \citep{Moon:2022}. It provides AUC estimation, CIs, and hypothesis testing. This test is equivalent to the Mann-Whitney U test, which assesses whether there is a significant difference in the distributions between the two groups. For example, the \fct{rss.AUC.test} can be used to test the null hypothesis of no difference in AUC between two groups ($H_0: AUC=\delta_0=0.5$):
\begin{verbatim}
R> rss.AUC.test(data1=rss.data1, data2=rss.data2, alpha=0.05, 
+  delta0=0.5)
$RSS_AUC
[1] 0.7123457

$CI
[1] 0.5586732 0.8156125

$`-2*log.LR`
[1] 6.63908

$p.value
[1] 0.009976544
\end{verbatim}

\subsection{Comparison with existing packages}

To our knowledge, the existing \proglang{R} packages that support RSS include \pkg{RSSampling}, \pkg{NSM3}, \pkg{RSStest}, and \pkg{RankedSetSampling}. While all these packages support RSS from population data using auxiliary variables for imperfect ranking - a standard feature in RSS methodologies - they vary in their scope and additional functionalities, with most focusing primarily on BRSS.

\pkg{RSSampling} provides sampling functions for both classical RSS and several modified RSS variants, such as Median RSS (MRSS), Percentile RSS (PRSS), Extreme RSS (ERSS), and Double RSS (DRSS). Additionally, it includes statistical inference methods for classical RSS, assuming balanced designs for both sampling and inference. \pkg{NSM3} includes only the classical RSS procedure as a sampling tool, along with critical value calculations for a nonparametric test. \pkg{RSStest} primarily focuses on mean testing for RSS and MRSS and generating RSS data under a normal distribution. Lastly, \pkg{RankedSetSampling} incorporates Judgment Post-Stratified Sampling (JPS) and RSS, offering sampling, mean estimation, and variance calculation, but it is limited to balanced designs.

The \pkg{generalRSS} package distinguishes itself by fully supporting both BRSS and URSS, offering advanced tools for flexible sampling and efficient sample allocation in URSS. Additionally, it includes parametric and nonparametric inference procedures for population means, medians, proportions, and AUC, along with simulation tools under predefined distributions (Normal, t, and Log-normal), leveraging a linear ranking model to incorporate imperfect ranking. Unlike other packages, it provides methods to calculate efficient sample allocations for URSS, improving estimation efficiency for both continuous and binary data.

For BRSS, the methods implemented in \pkg{generalRSS} align with the standard approaches available in other packages and are therefore not explicitly compared here. Instead, the focus is on the unique features of \pkg{generalRSS} for URSS and its broader applicability. Table~\ref{tab:comparison} summarizes the capabilities of \pkg{generalRSS} compared to other existing packages.

\begin{table}[!h]
	\centering
	\resizebox{.65\textheight}{!}{%
		\setlength{\tabcolsep}{6pt}%
		\renewcommand{\arraystretch}{1.05}%
		\begin{tabular}{P{2.5cm} P{3.3cm} P{3cm} P{3cm} P{3cm} P{3.3cm}}
			\hline
			\multirow{2}{*}{Feature} & \multicolumn{5}{c}{\proglang{R} package} \\ \cline{2-6}
			& \pkg{generalRSS} & \pkg{RSSampling} & \pkg{NSM3} & \pkg{RSStest} & \pkg{RankedSetSampling} \\ \hline
			
			\makecell[l]{Sampling} &
			\makecell[l]{Continuous\\ \li BRSS\\ \li URSS\\[2pt] Binary\\ \li BRSS\\ \li URSS} &
			\makecell[l]{Continuous\\ \li BRSS\\ \li MRSS\\ \li PRSS\\ \li ERSS\\ \li DRSS etc.} &
			\makecell[l]{Continuous\\ \li BRSS} &
			\makecell[l]{Continuous\\ \li BRSS\\ \li MRSS} &
			\makecell[l]{Continuous\\ \li BRSS}
			\\ \hline
			
			\makecell[l]{Simulation\\ Support} &
			\makecell[l]{Continuous\\ \li Normal\\ \li t\\ \li Log-normal\\[2pt] Binary\\ \li Binomial} &
			\makecell[l]{--} &
			\makecell[l]{--} &
			\makecell[l]{Continuous\\ \li Normal} &
			\makecell[l]{--}
			\\ \hline
			
			\makecell[l]{Sample\\ Allocation} &
			\makecell[l]{Continuous\\ \li Integer Neyman\\ \li Adjusted Neyman\\ \li LRC Allocations\\[2pt] Binary\\ \li Neyman} &
			\makecell[l]{--} &
			\makecell[l]{--} &
			\makecell[l]{--} &
			\makecell[l]{--}
			\\ \hline
			
			\makecell[l]{Parametric\\ Inference} &
			\makecell[l]{\li Mean Tests \\(BRSS, URSS)\\ \li Proportion Tests \\(BRSS, URSS)} &
			\makecell[l]{\li Mean Tests \\(BRSS)} &
			\makecell[l]{--} &
			\makecell[l]{\li Mean Tests \\(BRSS, MRSS)} &
			\makecell[l]{\li Mean Tests \\(BRSS)}
			\\ \hline
			
			\makecell[l]{Nonparametric\\ Inference} &
			\makecell[l]{\li Mean Tests \\(BRSS, URSS)\\ \li Median Tests \\(BRSS, URSS)\\ \li AUC Tests \\(BRSS, URSS)} &
			\makecell[l]{\li Median Tests \\(BRSS)} &
			\makecell[l]{\li Median \\Critical Values} &
			\makecell[l]{--} &
			\makecell[l]{--}
			\\ \hline
			
		\end{tabular}%
	} 
	\caption{Comparison of existing \proglang{R} packages for RSS.}
	\label{tab:comparison}
\end{table}

\section{Applications}
In this section, we consider two inference problems comparing two sampling methods: RSS and SRS. The first problem involves a one-sample mean inference, and the second involves a two-sample AUC inference. The goal is to estimate the population mean and AUC and compare the efficiency of RSS and SRS in terms of the CI length of each estimator. 

\subsection{One-sample problem}\label{sec:app-one}

We demonstrate the RSS sampling and inference procedures by estimating the mean body mass index (\code{BMI}) and testing the hypothesis $H_0: \mu=\mu_0$. The BMI is a widely used indicator of body fat based on weight and height, commonly employed to assess an individual's health status.

For this analysis, we use the US National Health and Nutrition Examination Survey (NHANES) dataset, available in the R package \pkg{NHANES} \citep{NHANES}. NHANES provides comprehensive health-related measurements collected from individuals across the United States. In this study, we focus on two key variables in the NHANES dataset: \code{BMI} as the outcome of interest ($Y$) and \code{weight} as the auxiliary variable ($X$) in RSS. 

The NHANES dataset originally contains BMI records for approximately 10,000 subjects. However, due to duplicated entries, only the first observation per subject is retained, resulting in a final dataset of 6,779 unique individuals. We treat this pre-processed dataset as the underlying population for our analysis.

\begin{figure}
    \centering
    \includegraphics[width=0.6\linewidth]{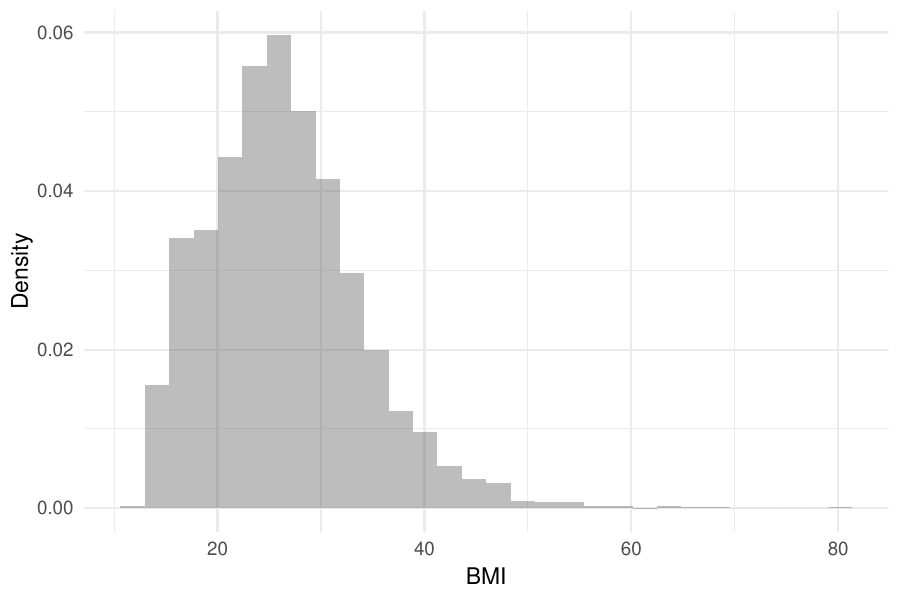}
    \caption{BMI distribution in the NHANES dataset \citep{NHANES}.}
    \label{fig:bmi_hist}
\end{figure}

We load the NHANES dataset from the R package \pkg{NHANES} as follows:
\begin{verbatim}
R> library("NHANES")
\end{verbatim}
The duplicated data are pre-processed by running:
\begin{verbatim}
R> library(dplyr)
R> dat = NHANES |> distinct(ID, .keep_all = TRUE)
R> mu0 = mean(BMI,na.rm=T)
R> mu0
[1] 26.48768
R> cor(na.omit(cbind(BMI, Weight)))
             BMI    Weight
BMI    1.0000000 0.9027414
Weight 0.9027414 1.0000000
\end{verbatim}
The true mean of \code{BMI} is 26.488 and the correlation between the auxiliary variable \code{Weight} and the outcome variable \code{BMI} is 0.903, indicating high-ranking quality. The histogram of the \code{BMI} in Figure~\ref{fig:bmi_hist} shows a skewed distribution, suggesting that URSS may be more efficient than SRS.

Using the NHANES data as the underlying population, we start with a BRSS design, using a set size of $H=3$ and a total sample size of $n=30$, resulting in a balanced allocation of $n_h=10$ for each stratum ($h=1, 2, 3$). However, missing data in the outcome variable can lead to URSS design. In such cases, we apply \fct{rss.design} to evaluate the efficiency of the initial sample allocation. If the allocation is deemed inefficient, adjusted allocations are calculated, and additional sampling is performed to obtain the final RSS data. Inference methods are then applied to the final RSS data. 

In practice, we first generate BRSS data using the auxiliary variable (\code{Weight}) to rank and select samples without measuring BMI. The outcome variable is then measured for the selected samples as described in Section~\ref{sec:rss}.
\begin{verbatim}
R> org.nsamp = c(10,10,10)
R> rss.dat = rss.sampling(ID=ID, X=Weight, H=3, nsamp=org.nsamp)
R> rss.dat$y = dat[ind,"BMI"]
R> head(rss.dat)
  rank    ID     y
1    1 64348 20.60
2    1 60341 17.67
3    1 58191    NA
4    1 53099 18.79
5    1 53171 29.85
6    1 51752 22.63
\end{verbatim}
Here, we observe missing values in the outcome \code{y}, resulting in a URSS sample allocation $(n_1=9, n_2=10, n_3=9)$ and total sample size $n=28$ as shown below:
\begin{verbatim}
R> rss.dat = na.omit(rss.dat)
R> org.nsamp = table(rss.dat$rank)
R> print(org.nsamp)
 1  2  3 
 9 10  9 
\end{verbatim}
Using this original URSS data, we test the null hypothesis $H_0:\mu=\mu_0$ using a t-test with \fct{rss.t.test} and the \code{sample} method.
\begin{verbatim}
R> org.t = rss.t.test(data1=rss.dat,mu0=mu0, method="sample")
R> print(org.t)
$RSS_mean
[1] 27.03478

$CI
[1] 24.91328 29.15628

$t
[1] 0.5383751

$df
[1] 19.74754

$p.value
[1] 0.5963351
R> diff(org.t$CI) 
\end{CodeInput}
\begin{CodeOutput}
[1] 4.242995
\end{verbatim}
In Figure~\ref{fig:bmi_hist}, the outcome variable shows a skewed distribution and URSS may offer improved efficiency, but the current sample allocation \code{org.nsamp} ($n_1=9, n_2=10, n_3=9$) is not guaranteed to be optimal. To address this, we use \fct{rss.design} to evaluate sample efficiency and adjust allocations accordingly:
\begin{verbatim}
R> alloc = RSS.design(data=rss.dat[,c("rank","y")])
R> add.samp = alloc$Adj.Neyman - org.nsamp
add.samp
1 2 3 
0 0 3
\end{verbatim}
We adopt the adjusted Neyman allocation \code{Adj.Neyman} and identify the need for three additional samples in the third stratum. From the remaining population except for the pre-selected samples, we sample again using the auxiliary variable without measuring the outcome. Then, we measure the outcome for the three newly selected samples:
\begin{verbatim}
R> after.dat = dat[-ind,]
R> add.dat = rss.sampling(ID=after.dat$ID, X=after.dat$Weight,
+  H=3, nsamp=add.samp)
R> add.dat$y = after.dat[add.ind,"BMI"]
\end{verbatim}
By merging the additional data with the original data, we obtain efficient URSS data.
\begin{verbatim}
R> update.dat = rbind(rss.dat, add.dat)
R> table(update.dat$rank) 
 1  2  3 
 9 10 12
\end{verbatim}
Using this updated RSS data, we reapply the t-test with \fct{rss.t.test}:
\begin{verbatim}
R> org.t = rss.t.test(data1=update.dat, mu0=mu0, method="sample")
R> print(org.t)
$RSS_mean
[1] 26.45867

$CI
[1] 24.46912 28.44822

$t
[1] -0.02996873

$df
[1] 26.12717

$p.value
[1] 0.9763197
R> diff(update.t$CI) 
[1] 3.979098
\end{verbatim}
When we compare the results before and after adding extra samples, we observe that the CI length decreases from 4.243 to 3.979 for RSS data.

We repeated the following procedure for 500 replicates to compare the performance of the original URSS, updated RSS, and SRS. For a fair comparison, SRS was assigned the same total sample size as the updated RSS. To generate the original URSS in each replicate, we introduced 10\% missing values into a BRSS dataset with $n_h=m = 10$, resulting in a URSS design. For the updated allocations, we selected the sample allocation provided by the \fct{rss.design} function that required the smallest number of additional samples. Table~\ref{tab:mucomparison} presents the average sample size, coverage probability, and 95\% CI over 500 replicates. Compared to the original URSS, the updated RSS achieves a significant reduction in CI length while maintaining a coverage probability close to the nominal 95\% level. Additionally, compared to SRS, the updated RSS demonstrates superior performance, achieving both a higher coverage probability and a shorter CI length. These findings underscore the efficiency of updated RSS in handling missing data and enhancing estimation precision.
\begin{table}[!ht]
\centering
\begin{tabular}{lccc}
\hline
Sampling & Sample size & Coverage Probability & CI Length \\ \hline
Origianl URSS     & 25.75 & 0.954                & 4.929                         \\
Updated RSS     & 27.10 & 0.954                & 4.665                         \\
SRS      & 27.10 & 0.944                & 5.933                        \\ \hline
\end{tabular}
\caption{Comparison of mean inference results from RSS and SRS over 500 replicates.}\label{tab:mucomparison}
\end{table}

\subsection{Two-sample problem}\label{sec:app-two}

We demonstrate the RSS sampling and inference procedures by comparing fasting plasma glucose (FPG) levels between two groups with and without diabetes through the estimation of the AUC ($\delta$). The FPG test determines whether blood sugar levels stay elevated after an extended fasting period, signaling potential issues with sugar metabolism \citep{american2022standards}. However, the FPG test may sometimes be inconvenient because it requires a fasting period of at least 8 hours. The HbA1c test, also called glycohemoglobin or hemoglobin A1c, is another valuable tool for diagnosing diabetes. It provides an average of blood sugar levels over the past 2-3 months and does not require any prior preparation, unlike the FPG test \citep{american2022standards}. In this application, we set the FPG as the outcome variable ($Y$) and glycohemoglobin as the auxiliary variable ($X$). 

AUC represents the probability that a randomly selected individual from the diabetes group has a higher FPG than a randomly selected individual from the non-diabetes group. For example, under $H_0: \delta=0.5$, the two distributions are identical, indicating no discriminatory ability between the groups based on FPG.

\begin{figure}
    \centering
    \includegraphics[width=0.6\linewidth]{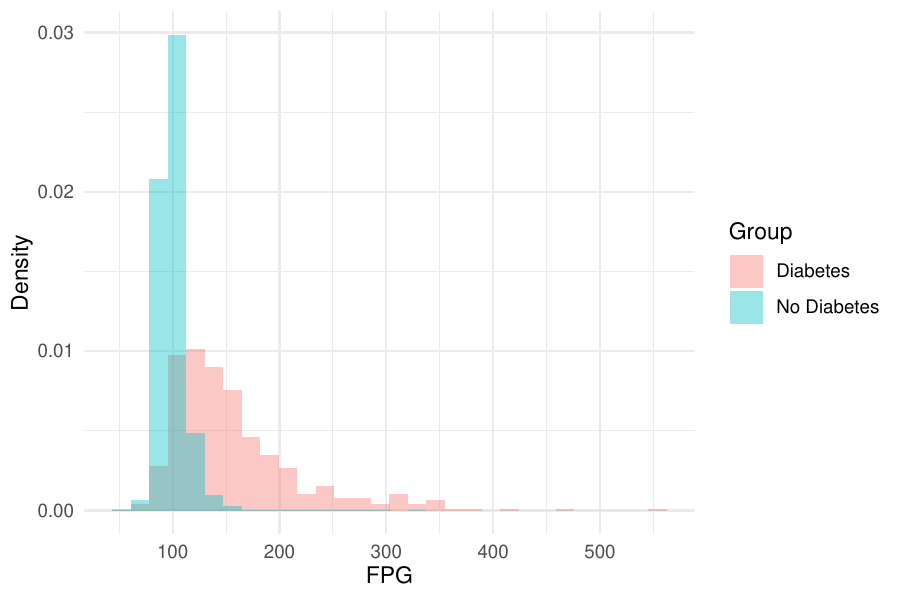}
    \caption{FPG distributions by diabetes status in the NHANES dataset \citep{centers2023national}.}
    \label{fig:hist-two}
\end{figure}

We analyze the NHANES dataset collected between 2021 and 2023, focusing on 3,540 subjects with data on FPG (\code{LBXGLU}), glycohemoglobin (\code{LBXGH}), and diabetes-related information (\code{DIQ010}) \citep{centers2023national}. 
We first separate the data into two groups by diabetes status.
\begin{verbatim}
R> grp1 = data |> filter(DIQ010 == 0) # non-diabetic
R> grp2 = data |> filter(DIQ010 == 1) # diabetic
R> mean(grp1$LBXGLU)
[1] 100.4777
R> mean(grp2$LBXGLU)
[1] 156.5644
R> cor(data$LBXGH, data$LBXGLU)
[1] 0.813826
\end{verbatim}
The true mean FPG levels are 100.478 for non-diabetes and 156.564 for diabetes populations, respectively. The histograms of FPG by diabetes status in Figure~\ref{fig:hist-two} highlight that the FPG distribution in the diabetes group is shifted towards higher values compared to the non-diabetes group. Also, the correlation between the outcome and auxiliary variables (i.e., FPG and glycohemoglobin) is 0.814, confirming strong ranking quality. 

Using \proglang{R} package \pkg{pROC}, we calculate the true AUC:
\begin{verbatim}
R> library(pROC)
R> delta0 = roc(data$DIQ010, data$LBXGLU, direction=c("<"), 
+  levels=c(0,1))$auc
R> delta0
Area under the curve: 0.8861
\end{verbatim}
In this section, we assume that the FPG level ($Y$) of the population is given, compared to Section~\ref{sec:app-one}. We use a set size of $H=3$ and a total sample size of $n=30$, a balanced allocation of $n_h=10$ for each stratum ($h=1, 2, 3$). Samples are ranked by the glycohemoglobin ($X$) and measure the outcome ($Y$) simultaneously for each group with a BRSS design:
\begin{verbatim}
R> org.nsamp1 = c(10,10,10)
R> org.nsamp2 = c(10,10,10)
R> H = length(org.nsamp1)
R> brss.grp1 = rss.sampling(ID=grp1$SEQN, X=grp1$LBXGH, 
+  Y=grp1$LBXGLU, H=H, nsamp=org.nsamp1)
R> brss.grp2 = rss.sampling(ID=grp2$SEQN, X=grp2$LBXGH, 
+  Y=grp2$LBXGLU, H=H, nsamp=org.nsamp2)
\end{verbatim}
We then apply \fct{rss.AUC.test} to these two RSS datasets to test the null hypothesis $H_0: AUC=0.8861$.
\begin{verbatim}
R> brss.auc = rss.AUC.test(data1=brss.grp1, data2=brss.grp2, 
+  alpha=0.05, delta0=delta0)
R> brss.auc
$RSS_AUC
[1] 0.9211111

$CI
[1] 0.8157051 0.9743021

$`-2*log.LR`
[1] 0.6119548

$p.value
[1] 0.4340528
\end{verbatim}
\fct{rss.AUC.test} returns the RSS AUC point estimate, CI, the ELR test statistics, and the $p$~value for the test.

We can improve the efficiency of RSS sampling using URSS design compared to BRSS. If AUC is expected to be high, the smallest rank group in the non-diabetic group and the largest rank group in the diabetic group provide limited information in AUC inference because the FPG levels are expected to be too small and too large, respectively. Therefore, a more precise inference is possible if we could have a larger number of samples from the largest rank group in the non-diabetic group and the larger number of samples from the smallest rank group in the diabetes group. To illustrate this, we set $(n_1=5, n_2=10, n_3=15)$ for the non-diabetic group and $(n_1=15, n_2=10, n_3=5)$ for the diabetic group. The URSS sampling and AUC inference can be done:
\begin{verbatim}
R> org.nsamp1 = c(5,10,15)
R> org.nsamp2 = c(15,10,5)
R> H = length(org.nsamp1)
R> urss.grp1 = rss.sampling(ID=grp1$SEQN, X=grp1$LBXGH, 
+  Y=grp1$LBXGLU, H=H, nsamp=org.nsamp1)
R> urss.grp2 = rss.sampling(ID=grp2$SEQN, X=grp2$LBXGH, 
+  Y=grp2$LBXGLU, H=H, nsamp=org.nsamp2)
R> urss.auc = rss.AUC.test(data1=urss.grp1, data2=urss.grp2,
+  alpha=0.05, delta0=delta0)
R> urss.auc
$RSS_AUC
[1] 0.8803704

$CI
[1] 0.7975823 0.9334440

$`-2*log.LR`
[1] 0.03003738

$p.value
[1] 0.8624054
\end{verbatim}
Using URSS, we estimate AUC as 0.880 with a CI length of 0.136, which is shorter compared to the CI length of 0.159 from BRSS data. 
\begin{verbatim}
R> diff(brss.auc$CI)
[1] 0.158597
R> diff(urss.auc$CI)
[1] 0.1358617
\end{verbatim}
We repeat the following procedure for 500 different samples and compare URSS and BRSS. We also compare the efficiency with SRS data. We sample two SRS data with the same total sample size of $n=30$ from the non-diabetes and the diabetes group, respectively, and estimate the AUC and its CI followed by \cite{Moon:2022}. Table~\ref{tab:AUCcomparison} shows the average coverage probability and the length of 95\% CI. URSS yields the highest coverage probability and the shortest CIs.
\begin{table}[!ht]
\centering
\begin{tabular}{lcc}
\hline
Sampling & Coverage Probability & CI Length \\ \hline
URSS     & 0.934                & 0.171                         \\
BRSS     & 0.930                & 0.175                         \\
SRS      & 0.930                & 0.181                         \\ \hline
\end{tabular}
\caption{Comparison of AUC inference results from RSS and SRS over 500 replicates.}\label{tab:AUCcomparison}
\end{table}
%


\section{Conclusion} \label{sec:con}
The \pkg{generalRSS} package provides a comprehensive framework for ranked set sampling (RSS), addressing challenges such as missing data, skewed population distributions, and binary data. Unlike traditional tools, it supports both balanced (BRSS) and unbalanced (URSS) designs, offering functions for flexible sampling, efficient sample allocation, and statistical inference for means, medians, proportions, and AUCs.

This paper compares \pkg{generalRSS} with existing RSS packages (Table~\ref{tab:comparison}) and demonstrates its application through real-world medical case studies. In the one-sample problem, an initial BRSS design became unbalanced due to missing data. By applying the \fct{rss.design} function, we optimized the URSS allocation, reducing CI lengths while maintaining coverage probability. In the two-sample problem, URSS was applied to AUC inference by designing the sample allocations based on the distributions of the two groups. Simulations confirmed that URSS improved estimation efficiency over both SRS and BRSS.

These findings highlight the advantages of unbalanced designs in RSS-based studies, particularly when missing data or skewed distributions limit traditional methods. The case studies illustrate the effectiveness of \pkg{generalRSS} in applied research and its broader applicability to inference problems beyond medical studies. Future work will focus on expanding inferential methods and refining sampling strategies to further enhance \pkg{generalRSS} as a tool for RSS-based analysis.


\section*{Computational details}

The results in this paper were obtained using \proglang{R}~4.3.1 with the following packages: \pkg{generalRSS}~0.1.3, \pkg{MASS}~7.3-60, \pkg{dplyr}~1.1.4, \pkg{ggplot2}~3.5.1, \pkg{pROC}~1.18.5, \pkg{tidyverse}~2.0.0, \pkg{haven}~2.5.4, \pkg{rootSolve}~1.8.2.4, \pkg{emplik}~1.3-1, and \pkg{NHANES}~2.1.0. \proglang{R} itself and all packages used are available from the Comprehensive \proglang{R} Archive Network (CRAN) at \url{https://CRAN.R-project.org/}.

All computations were carried out on a 64-bit Windows 11 Pro system (version 24H2) with a 13th Gen Intel(R) Core(TM) i7-13700 2.10GHz processor and 32 GB of RAM.

\section*{Acknowledgments}
This research was supported by the Learning \& Academic research institution for Master’s·PhD students, and Postdocs (LAMP) Program of the National Research Foundation of Korea (NRF) grant funded by the Ministry of Education (No. RS-2023-00285390) and Basic Science Research Program through the National Research Foundation of Korea (NRF) grant funded by the Ministry of Education (NRF-2021R1A6A1A10044950).


\bibliographystyle{plainnat}
\bibliography{RJ_generalRSS}


\end{document}